# THE TURING TEST FOR TELEPRESENCE

Mathias Johanson

Alkit Communications AB, Sallarängsbacken 2, Mölndal, Sweden

## ABSTRACT

*The quality of high-end videoconferencing systems has improved significantly over the last few years enabling a class of applications known as "telepresence" wherein the users engaged in a communication session experience a feeling of mutual presence in a shared virtual space. Telepresence systems have reached a maturity level that seriously challenges the old familiar truism that a face-to-face meeting is always better than a technology-mediated alternative. To explore the state of the art in telepresence technology and outline future opportunities, this paper proposes an optimality condition, expressed as a "Turing Test," whereby the subjective experience of using a telepresence system is compared to the corresponding face-to-face situation. The requirements and challenges of designing a system passing such a Turing Test for telepresence are analyzed with respect to the limits of human perception, and the feasibility of achieving this goal with currently available or near future technology is discussed.*

## 1. INTRODUCTION

In 1950, Alan Turing published a famous paper entitled "Computing machinery and intelligence" wherein he considers the question "Can machines think?" [1]. Although the scientific value of the paper can be questioned, it has undoubtedly been a great source of inspiration for many researchers within the field of artificial intelligence (AI). In search of an answer to the question, Turing proposes a reformulation whereby the original problem is reduced to an investigation of whether computers can successfully engage in an "imitation game." The game, as Turing describes it, is played by three participants: a man (A) a woman (B) and an interrogator (C) of any sex. The object of the game for the interrogator, who stays in a room apart from the other two, is to determine which of the other two is the man and which is the woman. The communication between the two rooms is by means of written notes only. The object of A in the game is to try to cause C to make the wrong identification, while B on the other hand should try to assist C making the right identification. The reformulated question posed by Turing is now, can a machine (i.e. a computer) be programmed to successfully play the part of A in the imitation game?

Although the imitation game is expressly presented as a challenge for the interrogator to determine the sex of the other two players, it is usually interpreted as the problem of deciding which of two subjects a computer is and which is a human.

Today, more than half a century since Turing posed his question; AI researchers devote little if any attention to the Turing Test as a measure of artificial intelligence. In fact, this was never the aim of Turing. His intention was to provide a tangible example to aid the discussion of the philosophy of artificial intelligence. However, the historical value of the test as an inspiration for AI researchers, computer scientists and technology developers to design systems and algorithms with powers comparable to human intelligence has been tremendous. In this sense, the Turing Test is often colloquially used to represent an ultimate goal of an AI system.

In this paper, I propose a test inspired by the Turing Test for evaluating the quality of telepresence systems. In place of Turing's question "Can machines think?" I pose the question "Can technology eliminate the effect of distance in interpersonal communication?". More simply put, is it possible to design a telepresence system that delivers a service that is indistinguishable from a face-to-face meeting? In a way, this is quite the opposite of Turing's test. In his imitation game, the subjects are allowed to communicate only by writing, so that the content of the conversation is in focus. In the telepresence test, we are on the contrary only interested in how the communication itself is experienced, not the subject matter of the conversation.

In order to study the question of whether technology can mediate interpersonal communication completely transparently, and to provide some sort of way to assess the quality attainable with a telepresence system, the problem must be defined more accurately. I will begin by defining what a telepresence system is.

## 2. TELEPRESENCE

The meaning of the term Telepresence in the context of this paper is the sensation of being physically present at the same location as another person (or any number of persons), although in reality the persons are at different locations. A Telepresence System is a set of technologies enabling this feeling of mutual presence by its users. Since vision and hearing are our main means of interpersonal communication, the key technologies of telepresence systems in this use of the term are typically concerned with audiovisual communication. A telepresence system can hence be seen as a high quality videoconferencing system. However, the term videoconferencing is heavily overloaded and can refer to almost any application of video-mediated communication. In order for a videoconferencing system to qualify as a telepresence system, there must be at least some ambition to present the appearance of the participants in a lifelike manner. A desktop videoconferencing system or a videophone application on a smartphone hence cannot be said to be a telepresence system.

The definition of telepresence discussed so far is focused on the sensation of being together with other persons who are not physically present at the same location, and being able to communicate with them. Another use of the term is concerned with the feeling of being physically present at a remote location (regardless of other people), or to have an effect on the remote environment, through telerobotics or teleoperation. Indeed, this was the focus of cognitive scientist Marvin Minsky's eponymous article in Omni Magazine in 1980, wherein the term was coined [2]. Today, the term Telepresence is most often used in the former meaning, i.e. in relation to high-end videoconferencing, partly due to the term being popularized as a marketing device by video-conferencing system vendors in the early 21st century. Telepresence in Minsky's connotation of the term is nowadays usually considered to fall within the field of Virtual Reality (VR).

### 2.1. Virtual Reality

The concepts of Telepresence and immersive VR are strongly interrelated. In an attempt to define VR in terms of human experience, rather than technological hardware, Jonathan Steuer [3] defines Telepresence as "…the experience of presence in an environment by means of a communication medium," where Presence is defined (based on Gibson [4]) "…as the sense of being in an environment." This is a generic definition covering both the communicative and immersive aspects of telepresence. The telepresence of interest here can be seen as a special case of the VR sense. For the rest of this paper, telepresence will refer to high end multimedia communication applications where the ambition is to give the users the illusion of being physically together in a shared space.

## 2.2. Immersion

Closely related to both telepresence and VR is the concept of immersion. In an immersive VR system, the user experiences himself or herself immersed in a virtual, computer generated world. This is typically achieved through head-mounted display technology, or through large, all-surrounding wall projection systems. Similarly, in an immersive telepresence system, the user experiences the technology mediating the presence of other users as being immersed transparently in the physical space he or she occupies. Whereas VR systems are mainly concerned with synthetic representations of environments and users (referred to as avatars), video-based telepresence systems rely on video capture and display technologies, carefully integrated in the physical environments of the users. Hybrid systems, using a combination of video and computer generated 3D graphics have also been designed. These approaches are often referred to as augmented reality or mixed reality systems. Telepresence systems relevant for our present study can include elements of both synthetic virtual environments and video communication technology.

## 2.3. Multi-user and multi-point systems

A salient feature of the telepresence systems of interest here is that they are by definition always multi-user systems. At a minimum two users, in a point-to-point communication session interconnecting two geographically separated sites, are required to qualify as a telepresence application. Telepresence sessions can also be truly multipoint, interconnecting more than two sites, and there can be more than one participant physically present at each site. It is furthermore required that all users — not just one or a subset of them — experience the sensation of telepresence. In a point-to-point two-user scenario, both users should experience the presence of the other user in a shared space. In that sense telepresence systems are by nature symmetrical.

## 2.4. Distance-spanning technology

In the above definition of telepresence I have stipulated, somewhat vaguely, that the users should be at different locations. To design a practical test in the spirit of Turing, I will have to be a bit more precise on what I mean by different locations. Is it enough for a system to interconnect two rooms in the same building to be considered a telepresence system? From a practical standpoint, a telepresence system must span distances that take some non-negligible time to travel to be of interest. Otherwise the users might as well get together physically for their meeting. For the present purposes, it suffices to require that the telepresence systems do not have a hard built-in limitation of the distance it can span. (There will be a hard upper limit imposed by the speed of light, as will be discussed later, but this is not a technological limitation.) This means that a system interconnecting two rooms in the same building is indeed considered a telepresence system, providing the technology the system is built on permits the same set-up to be realized between any two locations.

## 2.5 Multi-modality

Another issue is what modalities must be supported by a telepresence system. The simplistic answer to this question is that at least the modalities that are necessary to convey to the users a feeling of being physically co-located must be supported. In most practical situations, this means audio and video. There is no sense in claiming that an audio only communication channel is a telepresence system. Equally nonsensical in this respect is a video only system, making it impossible to communicate verbally. (Naturally, the users are assumed not be visually impaired or suffer from any other sensory disability.) Touch will generally not be required, since we can perfectly well have a feeling of someone else's presence without touching them. Smell will generally not either be required, although it might be conceivable that an olfactory communication system conveying the scent of someone's perfume might contribute to the overall feeling of presence. This effect should be marginal at best. I cannot foresee how any other sensory stimuli could contribute anything in this context.

It should go without saying that I will not consider any other means of experiencing presence of others than our traditional senses. It is interesting to note, however, that Turing in his seminal 1950 paper at some length discusses the possibilities of extra-sensory perception influencing his imitation game. Since this is probably the least flattering part of his work, I will not be concerned any further with it or by any other kind of metaphysical aspects of telepresence.

## 3. ASSESSING THE QUALITY OF TELEPRESENCE

The problem of assessing the quality of communication services has been studied for a long time, starting with audio quality measurements of telephony systems and subsequently video quality estimations of videoconferencing systems. When assessing the experienced quality of an audiovisual communication service, the available methods can be broadly classified as objective, subjective or hybrid approaches. Objective methods are based on statistical models for calculating how well a signal that is distorted by transmission over a noisy communication channel corresponds to the original signal. For instance, the Peak Signal to Noise Ratio (PSNR) is a very common objective quality metric for image and video communication services. Subjective methods, on the other hand, rely on having a panel of test subjects rate the perceived quality of media sequences in a controlled environment. Subjective quality is usually quantified by a Likert scale rank between one and five, representing the Mean Opinion Score (MOS) of the test panel. Hybrid methods, incorporating both objective and subjective elements, are typically based on utilizing some form of machine learning technique that is trained using subjective tests.

When referring to the quality of a telepresence system what is of concern is the quality of the subjective experience of presence provided by the system. Hence, to measure this quality, a subjective (or possibly hybrid) test is needed. Subjective media quality tests can be designed in many ways, some of which have been standardized, e.g. in ITU-T recommendations P.920 and P.1301. To make subjective tests comparable, standardized Absolute Category Rating (ACR) scales are used to score standardized test video sequences (or audio clips) under carefully controlled circumstances. To normalize the ratings, a Double Stimulus method is frequently used, which means that a reference signal is presented along with the assessment signal. The quality rating of a Double Stimulus test is expressed relative to the quality of the reference signal, which is known as Degradation Category Rating (DCR) or Comparative Category Rating (CCR). In CCR, the test subjects are not made aware of which signal is the assessment signal and which is the reference. This is also referred to as a Hidden Reference test.

It should be emphasized that the type of quality of interest here is not simply audio and video quality but the quality of the complete experience of presence provided by the system. Lately, the term Quality of Experience (QoE) has been popularized, denoting the subjective experience of a communication service. With this terminology, the aim of the present work is to device a QoE test for telepresence, which can be used to investigate whether an optimality criterion on the QoE test is achievable or not.

## 4. A TURING TEST FOR TELEPRESENCE

Based on the above definition of telepresence and the discussion about subjective tests of quality, I propose the following Turing Test for Telepresence (TTT):

A telepresence system is said to pass the Turing test if the experience of communicating with a remote user is subjectively indistinguishable from the experience of communicating with a locally present user.

This definition needs to be clarified in several ways to be possible to use as the basis for practical subjective tests, but the gist of the approach should be clear. What is meant by a telepresence system passing the TTT is simply that the system provides an audiovisual communication service that is as good as meeting a person face-to-face. In order to test whether a given system fulfills this optimality criterion, an experiment can be conducted wherein test subjects are invited to communicate with another test subject, both using the telepresence system under test and face-to-face. The test subjects will then be asked to rate the quality of the experience using an ACR scale. If the MOS of the quality of the telepresence system is as high as the face-to-face reference, the system could be said to have passed the TTT. In essence, this is a double stimulus test, possibly with a hidden reference, and as such it would be straightforward to design and even standardize. However, there are some practical issues that need to be addressed. First of all, the test subjects communicating with each other cannot be physically co-located and remote at the same time, so in order to carry out the reference test and the assessment test between the same two test subjects, at least one of them would have to travel between the test sessions. This is not only impractical, but undesirable, since it will impair the ability of the subjects to do a reliable comparative quality assessment. However, for our purposes, it is not strictly necessary that the remote and local test subject is the same person in both tests. What we are trying to measure is not the feeling of presence experienced by any particular subjects, but the general feeling of presence. In might thus suffice to arrange a test set-up wherein a subject (A) is presented with two other test subjects (B and C), one of whom (B) is physically present in the same room, whereas the other participant (C) is mediated by the system under test. If test subject A cannot tell if it is B or C who is the remote participant, the TTT is passed.
Figure 1 illustrates the set-up.

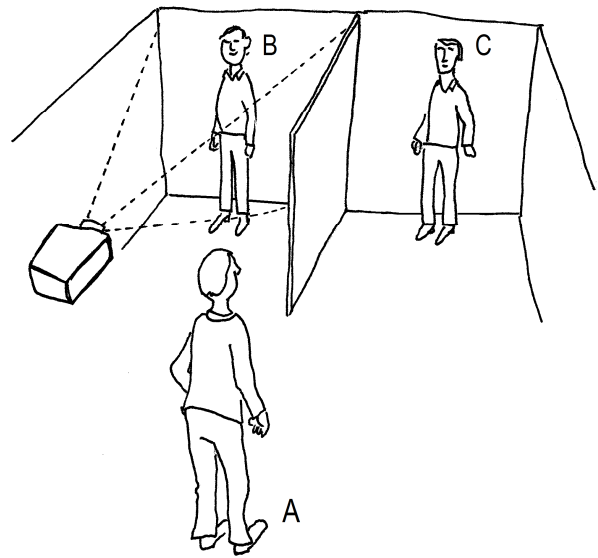

Figure 1. Subjective test for telepresence

As in any properly designed subjective test, one test run is of course not enough. There is in this case a 50-50 chance of guessing the correct answer. This is easily handled by arranging a sequence of independent tests, with different test subjects, recording the subject's opinion regarding which of the other participants is remote, and then confirming by statistical analysis whether the correct answer was overrepresented. The details of this are straightforward and need not be of further concern here. There are however other difficulties with this approach that must be addressed.

First of all, the definition of telepresence, as presented above, requires a symmetrical, bidirectional communication session, where both (or all) participants have to experience the feeling of presence. We must thus in the above-mentioned set-up also be able to test if C can tell whether A or a fourth participant D, physically present at C's location, is the remote participant. With this set-up, A and C are test subjects, while B and D are figurants, i.e. their subjective experience is not relevant to the test. Since the inclusion of figurants in subjective tests requires personnel and might negatively influence the outcome of the tests, it would be desirable with a test where all participants are test subjects. Such an approach is illustrated in Figure 2.

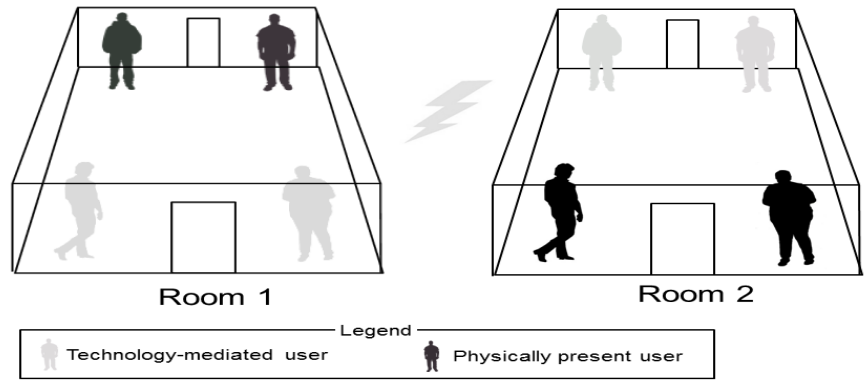

Figure 2. Symmetrical set-up for bidirectional test for telepresence

In the symmetrical test set-up of Figure 2, all four participants are test subjects. After the communication session each of them is questioned about who was the physically present participant in the meeting. Although it is an advantage to conduct a test with four test subjects simultaneously, in terms of time and people needed, it is still a quite complicated procedure, since the test subjects at the same location must be kept from meeting each other before the test. The requirements on the telepresence system are also significantly harder to meet compared to a baseline point-to-point telepresence test involving only two participants. For instance, the telepresence system of Figure 2 would have to support multiple consistent gaze directions.

A less demanding test set-up would be one wherein two test subjects (A and B) first are invited to communicate with each other using the telepresence system under test and then, after their interaction is finished, they are instructed to enter into another conversation with a physically present test subject (C and D respectively.) Then the telepresence system is used by C and D, whereupon each of the test subjects is questioned about which interaction was face-to-face and which was technology-mediated.

In this test, the remote and local participant are not available side-by-side for simultaneous scrutiny, but rather occur time sequentially. The set-up is much simpler from a practical standpoint, and probably an easier test for a telepresence system to pass. Nevertheless, it should be sufficiently challenging to serve as a baseline TTT.

Since geographically distributed tests might be complicated to administer, an alternative approach is to arrange the complete test set-up in the same building, with two rooms interconnected by the telepresence systems and a face-to-face meeting room for the reference test. In this case, the effect of transmission delay must be simulated, either by looping the network traffic through a reflector device at a remote location, or by buffering in a network emulator. The assessment test and the reference test can then be carried out one after the other, with the same two subjects.

Regardless of the technical set-up, there is the question of what means of interactions between the test subjects (and between test subjects and figurants, if present) should be allowed during the experiment. Naturally, the subjects cannot be allowed to move over to another subject and inspect by touch. The interaction should, without being unnecessarily restrictive, be limited to audiovisual communication, i.e. the subjects will be allowed to look at each other and speak to each other. Then the question arises whether the test subjects should be allowed unrestricted verbal communication, or if some sort of scripted conversation should be mandated. One can think of specific devices that could be invented by clever test subjects to assess specific aspects of the telepresence system, such as estimating the round-trip delay by instructing another participant to clap his or her hands immediately as he or she sees or hears a hand clap cue. (In a true distributed test, one test subject might simply ask another a question such as "where are you, I'm in Gothenburg" to reveal that they are not at the same place.)

A related question is whether the test should be blind or not, i.e. if the participants should be informed beforehand of the objective of the experiment. By studying these two questions in combination, it is easy to see that a blind freeform conversation test is the best choice. This means that the test subjects are not informed before the test what the goal of the test is. They are merely instructed to enter into conversation with the other parties. In an interview after the test, the test subjects are informed about the reason for the test and questioned about whom of the participants was experienced as being remote and who was locally present. This encourages a natural interaction between the test subjects.

Another question is how close to each other (spatially) the test subjects should be allowed to get in the combined physical and virtual space. Naturally, there is a big difference distinguishing between a mediated remote and a local user at viewing distance of ten meters, compared to an arms-length of separation. The sensible distance to prescribe for the TTT is a comfortable face-to-face conversation distance, which typically is between one and a couple of meters depending on the meeting situation. Appreciating the importance of this parameter, I will for now allow for a bit of leeway, accommodating for different application scopes.

There is also a question of whether full-body visual representations should be mandated, or if, for instance, upper-body only scenarios should also be allowed. The latter could be a very suitable TTT for business-meeting telepresence systems, where a round-table meeting is the typical application scenario. Again, I will not bother to prescribe the specific details, concentrating instead on the conceptual approach, since my goal is not really to devise a practical test but to explore whether a TTT designed along the basic lines described here can be passed by telepresence systems available today or in the near future. To do this, I will again turn to a discussion about QoE and explore the individual quality dimensions accounting for the total experience of presence.

## 4. QUALITY REQUIREMENTS OF TELEPRESENCE

The QoE of a telepresence system is directly dependent on the representational richness of the modalities supported and the ability of its users to interact with each other in a transparent and intuitive way.

As discussed above, the modalities (i.e. the sense perceptions) that must at minimum be supported for a system to pass the TTT are vision and hearing, so we can categorize our quality requirements into video requirements, audio requirements, and cross-modality requirements involving both audio and video (such as lip sync). The interactional or conversational requirements include the implications of communication delay on QoE and higher level conversational aspects such as eye-contact and gaze directions.

It is furthermore required that a telepresence system should present a shared environment for all participants which is consistent with the users' expectations and experiences. For instance, this includes consistent lighting conditions so that shadows are consistent with the room's light sources. Another example might be that the voice of a user should sound like the voice of this particular user; it is not enough that the audio quality is excellent and perfectly synchronized with the lip movements of the user.

To investigate what the prospects are of current and near-future telepresence systems to pass the TTT, I will analyze the quality requirements identified in some detail with respect to the limits of human perception. Although the intention is not to suggest technological solutions for how to best design a system passing the TTT, there will be some need for speculation about feasible solutions, in order to assess whether the current state of the art in a particular technology area is sufficiently advanced for the test to be passed.

## 5.1. Video quality requirements

Fundamentally, if a telepresence system cannot capture and reproduce high enough quality video signals, the TTT cannot be passed. Thus, the performance of the input and output video devices — typically cameras and displays or projection systems — is critical to the overall quality. Indirectly, signal processing and communication system performance also influence the outcome: if a signal of high enough quality cannot be processed and transmitted over a communication channel in real time, the TTT will fail.

The quality dimensions of visual perception include spatial resolution, temporal resolution (accommodating for motion perception and other changes in the field of vision), stereoacuity (accounting for binocular depth-perception), color fidelity (discrimination of different hues and saturations of colors) and dynamic range (perceiving different levels of luminance or brightness). From these we can derive three main quality parameters determining the quality delivered by video capture and presentation devices: the pixel density of the devices (determining the spatial resolution achievable), the frame rate (determining the temporal resolution) and the precision used to represent each pixel (or sub-pixel, since each pixel is composed of a red, a green and a blue component).

The question is then, what resolution, what frame rate and how many bits per pixel must the input and output devices of a telepresence system support in order to achieve high enough video quality to pass the TTT? And to what extent can technology available today meet these requirements?

### 5.1.1. Video resolution

There are two main factors limiting the spatial resolution perceivable by the human visual system: the eye's optical properties and foveal cone spacing. Subjective tests as well as studies of photoreceptor density in human eyes have shown that the human visual system can resolve spatial detail down to a visual angle of approximately half an arc-minute (i.e. 1/120 degree) [5, 6]. With a 2 m viewing distance (consistent with our loosely defined "comfortable face-to-face conversation distance" above), this gives a maximum pixel size of 2 * $tan$(1/120) m, i.e. approximately 0.3 mm. For a display big enough to render a full scale person, let's say 2 by 0.5 m, this gives a required resolution of 2000/0.3 by 500/0.3, or about 6700 by 1700 (i.e. about 11 megapixels). With 8k Ultra High Definition Television displays supporting 7680 × 4320 resolution being already commercially available, an 80 inch UHDTV would seem to easily provide the spatial resolution needed to pass the TTT.

It must be remembered that this resolution requirement is derived from estimates of the theoretical limits of the human visual system's ability to resolve fine detail. Psychovisual experiments are typically performed by letting test subjects view high frequency gratings, to determine the limits of their resolving power. In a real telepresence situation, the images presented to users are typically of much lower spatial frequency and thus the resolution requirements will be lower, possibly as low as 1920x1080, the common HD format supported by many off-the-shelf videoconferencing products. Based on this, I think we can safely assume that the QoE achievable by contemporary telepresence systems is not limited by spatial resolution in itself. There is however another characteristic of the human visual system that might influence the resolution required, namely the ability to perceive three-dimensional depth through binocular disparity.

### 5.1.2. Stereoscopic video requirements

The fact that we have two eyes implies that the brain constantly processes two slightly different visual signals, which it fuses into one three dimensional view. To reproduce this in a telepresence system, each user should be presented with two slightly different video signals, one for each eye. The two signals furthermore depend on the position and movement of the viewer's head. The traditional way of achieving true three dimensional visualization is to use head-mounted displays or view-filtering stereo glasses to present different images to the left and right eye. However, for telepresence applications this is typically not feasible, since the eyewear will, at least with technology available today, obscure the face of the user, making mediated face-to-face communication compatible with our definition of telepresence impossible. More recently, however, autostereoscopic display technology has appeared, enabling stereoscopic visualization without the need for eyewear. Such displays employ various techniques such as parallax barriers or lenticular lenses to present different images for the left and right eyes, relying on the different viewing positions of the eyes. Multiview autostereoscopic displays, providing more than two visual channels, not only enable parallax-based depth perception, but also to some extent support the change in perspective resulting from head motion. For instance, by moving your head to the side, you are able to see more of an object that is partly occluded by another object in front of it. In order for this effect to be realistically reproduced by a multiview autostereoscopic display, the number of visual channels supported must be fairly large, particularly for large head movements. Going back to my discussion about the resolution requirements, each channel of a multiview display will theoretically require the UHDTV resolution, although as previously discussed, in practice the required resolution is probably considerably lower. Nevertheless, with the use of multiview autostereoscopic displays, the resolution requirements will increase linearly with the number of visual channels supported by the device. Even for a moderate number of channels required, this will indeed present a challenge beyond the current state of the art in multiview autostereoscopic display technology. To further analyze this requirement, I will estimate the number of visual channels needed. First however, I will explore in some more detail whether stereoscopic visualization is really necessary for telepresence applications.

### 5.1.3. The need for stereoscopic video

Since a frontal view of a person (i.e. a traditional face-to-face conversation view, as in a typical telepresence situation) is in fact rather flat, it is questionable whether parallax-based depth cues are at all necessary to render a sufficiently realistic view. Furthermore, it is a well-known fact that much of our depth perception come from monoscopic cues, like interposition and relative size, so the need for binocular depth is not at all obvious. Unless the user extends an arm or otherwise exaggerates the depth, the typical relative depth in the frontal view of a person is limited to a few centimeters at most, representing for instance the distance from the tip of the nose to the ear. If we assume this distance to be 10 cm and the interlocutor distance to be 2 m as before, we can calculate the stereoscopic visual acuity (or stereoacuity) necessary to perceive this relative depth.

Stereoacuity is defined by the minimum perceivable angle of binocular disparity, i.e., the difference in binocular parallax between the two eyes. It can be calculated using the formula $a = x * d / z^2$, where $a$ is the angular disparity acuity, $x$ is the interocular separation of the observer, $d$ is the relative depth and $z$ is the viewing distance (with $z$ and $d$ parallel). With $d = 0.1$, $z = 2.0$ and $x = 0.065$ (representing an average eye separation of 6.5 cm) we get $a = 0.001625$, in radians, which translates to about 5.6 arc minutes. This is about ten times the typical stereoacuity under favorable conditions, established experimentally and approximately constant over the distance range of interest here [7], so depths of merely around 1 cm should in fact be discernible, from binocular stereopsis alone at the 2 m viewing distance. This means that even facial features (e.g. nasal protrusion, orbital cavity) contain enough depth information to be perceptible, and hence stereoscopic vision seems in fact to be highly relevant in telepresence.

### 5.1.4. Multiview auto stereoscopic display resolution

Appreciating that stereoscopic presentation will probably be necessary, I will now return to the question of how many visual channels will be needed for an autostereoscopic display. If we assume a maximum horizontal head motion in a telepresence situation to be about 20 cm (i.e. a head displacement of 10 cm to the left and right of the normal position respectively) and the vertical head motion to be negligible (since you typically do not move your head up and down much in a conversational situation), this corresponds to a visual angle of $2*arctan(0.1/2)$ at the 2 m viewing distance, or about 5.7 degrees. Dividing this value by the visual acuity of half a minute of arc cited above (which conveniently enough happens to be about the same for stereoacuity), we get 687 as the number of visual channels needed to support head motion with perceptually unnoticeable precision, both with regard to stereoscopic and monoscopic cues. Again, since the visual acuity limit is determined from high-frequency gratings, and the typical visual scene of our application can be assumed to be of much lower frequencies, this is probably an overestimate, but can nevertheless serve as a reasonable upper limit.

### 5.1.5. Multiview video channel generation

Whereas our initial resolution requirement for monoscopic vision seemed relatively tractable, the aggregate resolution requirement for stereoscopic vision supporting head motion of approximately 7.6 gigapixels (687 * 11 megapixels) seems very challenging indeed, both from a display and a video capture perspective. If we briefly consider the capture side of the system, a camera array of 687 image sensors, properly aligned to generate the required 5.7 degree field of view, would ideally be required. With currently available technology this seems quite unattainable, which might suggest that another approach is needed, relying instead on head tracking and selective generation of the visual channels. Since the user only sees two visual channels at a time (one with each eye), it should be possible, at least theoretically, to generate the two video signals based on the current head position, either by synthetic view generation or by signaling the head position to the transmitting side to move the two (or more) image sensors of the capture system. The latter approach suffers from the problem of requiring a round-trip transmission delay to adjust the viewpoint in response to head motion, which might be noticeable to the user as a visual lag when moving the head. A possible remedy for this might be to transmit not only two but a group of channels surrounding the two viewpoints of the currently tracked head position. The number of channels thus required to be transmitted simultaneously would be determined by the maximum speed of head motion and the round trip delay of the communication channel. If we reasonably assume the maximum speed of head motion to be 1 m/s and the round trip delay to be 10 ms (for reasons that will be explained later), this means that the viewpoint moves at most 1 cm in one round-trip, corresponding to 20/687 or about 34 visual channels. Lowering the required number of visual channels by a factor 20 this way of course seems favorable, but the technological challenges of the suggested tracking-based system should not be underestimated. This approach furthermore only supports one viewer at each site.

The other alternative suggested above, i.e. to instead rely on synthetic view generation of the visual channels, is a fundamentally different approach akin to how views are rendered in VR systems. With this approach, the users of the telepresence system are represented by 3D models rendered stereoscopically using conventional 3D computer graphics visualization techniques. Considering the tremendous success of 3D graphics for video games and animations in motion pictures, this may at first seem like a straightforward application of readily available technology, but unlike these applications, a telepresence system must generate the models of remote participants in real time using a 3D scanner. Many types of 3D scanners are available, including laser-based devices that generate a point cloud of depth measurements from which a 3D model of the subject can be constructed. Alternatively, multiple camera systems generate a depth map by matching corresponding pixels in the different camera views and calculating the disparity. Structured light 3D scanners project a pattern on the subject and measure the geometric deformation of the pattern in images captured by a camera from a different angle than the projector. For the application of interest here, the constructed 3D model must be rendered using texture mapping, with textures generated in real time from cameras. The main challenge involved in this type of augmented or mixed reality system is to achieve low enough latency and high enough processing performance in order to deliver high enough frame rates for motion in the rendered 3D scene to appear smooth. Despite rapid progress in this field, such processing must be considered beyond the limits of contemporary VR technology.

### 5.1.6. Frame rate

Regardless of how images of remote participants are generated in a telepresence system (i.e. synthetically using VR rendering techniques or live capture from one or more video cameras, or using a hybrid approach), there is the question of what frame rate is needed for capture and display to avoid perceptible flicker and to give an impression of smooth and natural motion. Flicker is a problem with displays of a type that flash each image onto a screen, followed by a fading interval until the next image is displayed. If the refresh rate is not high enough, the inter-frame fading will be noticed as flicker. Cathode ray tube (CRT) and digital micromirror based devices are prone to this effect. The flicker fusion threshold, which is the frequency at which a flashing light stimulus appears to be completely steady to an average human observer, depends on a number of parameters such as the intensity of the stimulus. Its maximum value, at high luminance, has been determined experimentally to be about 50 to 60 Hz [8]. This, together with the fact that the mains frequencies happen to be 50 and 60 Hz respectively in Europe and the US, is the main reason why television standards have chosen 50 or 60 Hz refresh rates (equivalent to 25 and 30 interlaced frames per second). Avoiding flicker is also the reason why traditional movie projectors display each of the 24 frames per second (fps) twice or thrice. Unfortunately, this refresh rate has often been erroneously believed to be the limiting rate for perception of motion in a scene [9]. Subjective tests show that the perceptual limiting frame rate for blur and jerkiness related to motion is in fact as high as about 250 Hz [9, 10]. In the psychophysical experiments determining this limit, high motion video clips are shown to a test panel at different frame rates for subjective evaluation. Telepresence scenes, on the other hand, are characterized by low motion, like lip-movement, head movements, posture changes, waving of hands and similar. There will be no camera pans and zooms, which is otherwise a common source for scene motion. To estimate the frame rate necessary to realistically present the motion of a telepresence situation, we first note that humans perceive moving objects through an eye movement tracking mechanism known as smooth pursuit, whereby the image of the moving target is kept on or near the fovea. The maximum velocity of objects that can be tracked by the eye in this way has been experimentally determined to be around 100 degrees per second [11]. Combined with the 250 Hz limiting frame rate, this implies that a movement of around 0.4 degrees per frame can be tolerated for perceptibly smooth motion. If we reasonably assume that the maximum velocity of moving objects (heads, hands, lips) in a telepresence situation is around 1 m/s, which translates to an angular speed of 26 degrees per second at a 2 m viewing distance, this gives a frame rate requirement of 26/0.4 = 65 fps. Since this is also above the flicker threshold, it might serve as an indication of the frame rate required to pass the TTT.

### 5.1.7. Pixel precision (bits-per-pixel)

I have now at some length analyzed the resolution and frame rate requirements of telepresence. The third of the basic visual quality parameters to be discussed is the precision needed to represent each pixel. This parameter influences the number of colors that can be represented as well as the dynamic range (i.e. the contrast ratio) achievable. Perhaps somewhat surprisingly, there is a lot of confusion about how many colors the human visual system can discern, with estimates ranging from 100,000 to 10 million [12]. The source of this confusion is probably related to the fact that the visual system does not directly process the tristimulus values of the photoreceptors of the eye. Rather, the brain translates from the retinal trichromatic representation into an opponent processing model based on two color-difference (chrominance) signals and one luminance signal. The wide difference in the estimated numbers of perceivable colors is thus dependent on whether the number corresponds to the full range of tristimulus values (which would likely be in the range one to ten millions), or to the number of chrominance levels, normalized by luminance (which would probably be in the order of 100,000). In digital imaging, 8 bits are usually used to represent each component (red, green and blue) in the RGB colorspace, which gives a full range of $2^{24}$ (about 16 million) different tristimulus levels. Correspondingly, in the opponent-processing model, transforming from the RGB colorspace to a luminance/chrominance colorspace (YUV) the way it is usually done in digital image processing, gives 16 bits of chrominance and 8 bits of luminance. This translates to $2^{16} = 65536$ distinct chrominance levels (hues) at 256 luminance (intensity) levels. Consequently, the 24 bit pixel precision seems to be sufficient from the perspective of the trichromatic model, whereas the opponent-processing model suggests this might be slightly below the limits of the human visual system. Increasing the number of bits per component to 10, which is common in digital video processing, increases the chroma levels to just above one million and the luminance levels to 1024. The latter would furthermore require a display technology capable of roughly 1000:1 contrast ratios, which is well within the limits of current technology. Although the full dynamic range of the human visual system is way beyond this (particularly if scotopic and mesopic vision is taken into account) the process of light adaptation within the retina in practice limits the useful dynamic range to about 100:1 [13]. This strongly suggests that 30 bits per pixel should be more than enough for most applications, including telepresence. Considering the fact that the spectral range of telepresence scenes is rather limited, 24 bits per pixels can be assumed to give high enough visual fidelity to pass the TTT.

### 5.1.8. Eye Contact and gaze awareness

Eye contact is well-known to be problematic in videoconferencing, due to the fact that the camera is usually not placed directly in the optical path between the user's eyes and the display showing the remote user. Figure 3 illustrates the problem. Since eye contact is a very important social cue in face-to-face communication, it will be crucial for a telepresence system not to compromise the perception of eye contact.

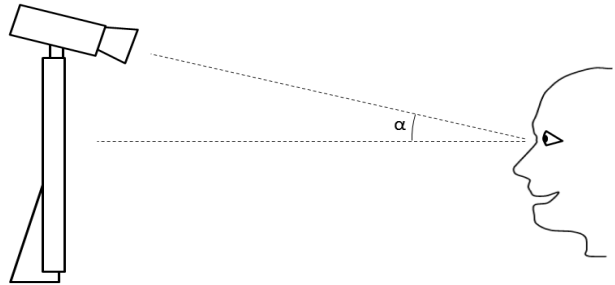

Figure 3. Classic eye contact problem in videoconferencing.

To estimate the human sensitivity to eye contact in telepresence, we note that the perception of eye contact depends on the ability to detect in which direction a remote participant's eyes are pointing, which in turn depends on the rotation of the eyeballs within the eye sockets. The rotation of the eye causes a change in the position of the iris within the sclera. (The pupil is always centered in the iris, so the displacements of the iris and the pupil are the same.) As estimated above, a visual acuity of half an arc minute corresponds to a spatial displacement of about 0.3 mm at a 2 meter viewing distance. If we assume an eyeball to have a diameter of approximately 20 millimeters, a just noticeable rotation of the eyeball at our 2 meter viewing distance will be *arctan*(0.3/20) or about 0.85 degrees. This estimate is in agreement with a large number of subjective tests of eye contact sensitivity, reporting approximately one degree as the limit [14, 15, 16]. Furthermore, subjective tests have shown sensitivity to eye contact to be asymmetric, with considerably lower sensitivity in the vertical direction (up to about 5 degrees). This means that if a camera is not put directly in the optical path between the user and the display, it should be placed above the display (as in Figure 3).

A one degree sensitivity corresponds to a maximum tolerable displacement of the camera from the optical path of about 3.5 cm at the 2 meter viewing distance. The corresponding number for a 5 degree tolerance is about 17 cm, indicating that camera placement strategies for achieving eye contact in telepresence do not necessarily have to rely on inserting cameras directly in the optical path by means of semi-transparent mirrors or similar.

Telepresence sessions involving more than two participants are particularly challenging, since eye contact must be possible between every pair of participants, and the gaze directions must at all times be consistent. For instance, in a telepresence session involving three participants A, B and C, when A and B looks at each other, not only need A and B perceive eye contact, C must also perceive that they are looking at each other. When A shifts gaze direction to C, A and C must perceive eye contact and B must perceive that A and C are looking at each other.

Although there are technological challenges in supporting consistent gaze directions and eye contact, particularly in multipoint, multi-user situations, many approaches have been suggested and implemented, indicating that solutions good enough to pass the TTT are feasible.

### 5.1.9. Other video requirements

In addition to the basic requirements in terms of resolution, frame rate and pixel precision analyzed above, and the more socially oriented cues of eye contact and gaze awareness, there are also a number of additional aspects of video quality that will be required to meet the TTT. There must be consistent lighting conditions so that for instance shadows are consistent with the room's light sources and the color temperature is uniform. There must of course be no noticeable rendering distortions such as moiré effects, light reflections or other tell-tale signs that a remote participant is not actually present but video-mediated. Depth cues must be consistent between renderings of remote participants and the physical room, and the depth perception must be realistic in the presence of head motion. Perspectives and relative size of objects must be consistent and realistic. Specifically, the users must be rendered at their true sizes or sufficiently close to, not to be perceptibly anomalous.

## 5.2. Audio quality requirements

Equally important as the quality of video signals, in order for a system to pass the TTT, is the quality of audio signals. If the voice of a remote participant is not faithfully conveyed by the system, the illusion of presence is lost. Not only voice but also ambient sounds need to be mediated faithfully by the system, including any sound that can be present in an interpersonal communication session (e.g. snapping of fingers, sighing, coughing, etc.) There must be no discernible distortions, echoes or other anomalies. The audio communication must be truly duplex, making interruptions possible, as in a face-to-face conversation. The audio characteristics of the speech signals of remote participants must fit the acoustics of the room where they are played back. Volume levels must be realistic and directional audio cues must be consistent with the visual information.

The two fundamental quality aspects of audio signals are the sampling rate (deciding the frequency range supported) and the sample precision (deciding the dynamic range, or signal-to-noise ratio).

### 5.2.1. Sampling rate

A healthy young person can hear frequencies from about 20 Hz up to about 20 kHz. Consequently, the maximum audio frequency range that a telepresence system needs to support is about 20 kHz. It follows from the Nyquist-Shannon sampling theorem that a sampling rate of at least 40 kHz is required for a digital communication system to be able to convey the full frequency range. Digital music recording systems usually operate at 44.1 kHz or 48 kHz, so this is no problem with state of the art technology.

### 5.2.2. Sample precision (bits-per-sample)

An upper limit on the number of bits per audio sample required is given by the dynamic range of the human auditory system, which is roughly 140 dB [17]. Speech, however, which is the audio signals of main concern in telepresence, is normally perceived over a dynamic range of about 40 dB [18]. The 16-bit sample precision used for compact disc (CD) recordings has a theoretical dynamic range of about 96 dB [19], and could hence be expected to be sufficient. State of the art audio recording and playback technology can without problem support 24 bits per sample, with theoretical dynamic range of 144 dB. Although the audio recording and playback chain necessarily includes analog circuitry significantly limiting the practically achievable dynamic range, it must nevertheless be considered plausible that audio quality in terms of frequency response and dynamic range sufficient to pass the TTT should be achievable with state of the art audio technology. There are however other aspects of audio quality that need to be considered.

### 5.2.3. Spatial localization

Our ability to localize sound sources in space and to separate sounds based on their spatial locations relies on both monaural and binaural cues, with the latter providing the best precision. The two main binaural cues are differences in phase and intensity of the received signal in the two ears. Psychoacoustic experiments have shown that a sound source can be localized with a precision as high as about one degree in the horizontal direction, and considerably worse for vertical localization [20]. A one degree angle at a distance of two meters (our nominal telepresence interlocutor distance) corresponds to about 3.5 cm, indicating that in order for the sound of a mediated participant's voice to be perceived as coming from the right direction, it should be played back by a loudspeaker very close to where the participant's mouth is rendered. However, since the visual and aural cues of voice perception are strongly interrelated, the availability of a visual positioning cue can be expected to override the aural cue for positioning, providing a bit of leeway in the positioning of speakers. Moreover, with stereophonic or multichannel audio systems, a skilled sound engineer can position audio sources in space with a precision that can be expected to be good enough to pass the TTT.

### 5.2.4. Acoustic echoes and consistent room acoustics

A well-known source of problems in traditional videoconferencing and telepresence is echoes resulting when the audio signal from the speaker is picked up by the microphone and transmitted back to the originator. To avoid this, echo-cancelling devices or software is used which filter out the playback signal from the input microphone signal. Since the recorded (echo) signal is not exactly the same as the playback signal, a transfer function is computed, which models the change the signal undergoes when passing from speaker to microphone. Although echo cancelling technology has improved considerably over the last few years, it is difficult to do perfectly, since the transfer function is dependent on the acoustics of the room, which changes slightly when the speaker changes position.

Perfectly imperceptible acoustic echo cancellation, as required by a telepresence system to pass the TTT, must still be considered a significant challenge.

Another issue related to acoustics is the sound characteristics of the combined physical and virtual space of the telepresence system. The subjective perception of the acoustics of the environment, sometimes referred to as "ambience", including background noise, reverberation and acoustic resonance, must be consistent with a physically co-located conversation.

### 5.3. Cross-modality requirements

A critical requirement for a telepresence system to pass the TTT is that all media signals are presented in a time synchronized fashion with a precision determined by the limits of human perception. Specifically, audio and video signals of a speaker must be synchronized so that the movements of the lips appear consistent with the voice signal.

Subjective tests of human sensitivity to lip synchronization in television scenarios have revealed the threshold of detectability to be about 40 ms for audio leading video and about 60 ms for video leading audio [21]. The reason for the asymmetry is most likely that we in real life never experience audio signals leading video, whereas the reverse case is what happens when increasing the distance between interlocutors, due to the huge difference in propagation delay of sound and light. This fact furthermore demonstrates that perceived media synchronization does not depend on sound caused by an event reaching the listener's ears at the same time as the light from the image depicting that action. The audio signal is always behind the video with the magnitude of the latency depending on the distance. In the prototypical telepresence set-up with at two meter separation between two participants, the audio signal should trail the video by about 6 milliseconds to appear perfectly synchronized. The interval of imperceptibility, from -40 to 60 milliseconds, gives some latitude, but a telepresence system must be designed with great attention to synchronization of media streams or the sensation of presence will be effectively lost.

Since video processing is typically much more computationally complex than audio processing, the standard practice in videoconferencing systems is to delay the audio signal at the receiver side to match the delay of the video signal at presentation time. This introduces end-to-end audio latency which has implications on acoustic echo suppression as discussed above and on overall latency issues affecting the interactivity of conversation, discussed next.

## 5.4. Conversational and interactional quality

The delay imposed by signal processing and transmission in a telepresence system must be imperceptible during conversation in order for the TTT to be passed. If a substantial round-trip delay is introduced by the telepresence system, this will be noticed by the users by responses to utterances being slower than expected. A related effect will be that two persons start talking at the same time, a phenomenon sometimes referred to as "double talk."

The problems that can appear are best illustrated by an example. Consider a time-delayed telepresence conversation between two people, A and B where $t$ is the total one-way delay added by the system. Assume that A asks B a question that can be expected to result in a prompt response. The question will arrive at B after the time $t$, whereupon B's response will require the same propagation time and will hence arrive at A at time $2t$. (We ignore the reaction time of B for now.) Meanwhile, since A expected a prompt reply, he or she may assume that B didn't hear the question and will repeat it. This may lead to double-talk if the response from B arrives while A is repeating the question. Depending on the length of B's reply, this might cause B to stop talking, assuming A is interrupting the response. Disorder in turn-taking may thus result, possibly with the communication breaking down altogether if $t$ is sufficiently large.

### 5.4.1. The implications of delay in telepresence

It should be pointed out that the only way for users to directly detect delay in a telepresence scenario is in transitions between one speaker and another; there is no way to detect delay if only one user is speaking. (Indirectly, the users may infer that there is delay due to imperfections in acoustic echo cancellation, or similar shortcomings.) While this might suggest that adding a limited delay to conversations should not be too disruptive, we must also keep in mind that turn-taking in verbal communication is a task performed very efficiently by the human cognitive system without imposing much subjectively perceived effort – it is a complex skill we are very good at, and one can therefore hypothesize that even slight deviations from a normal conversational pattern will be detectable.

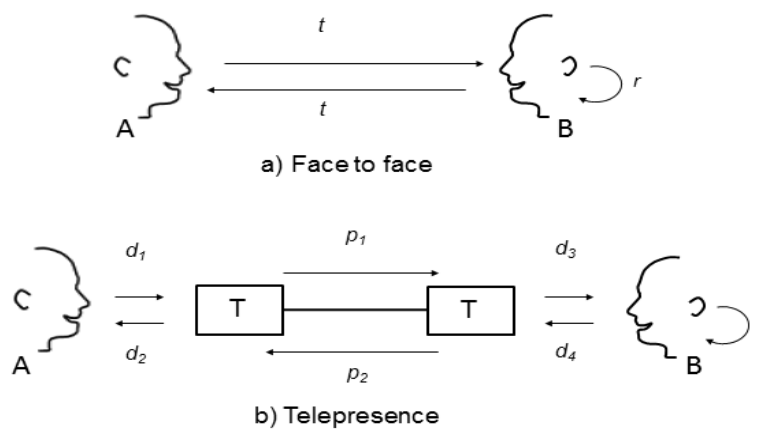

Figure 4.  Sources of delay in face-to-face and mediated conversations.

In a face-to-face conversation, the perceived response time to verbal utterances is determined by the propagation delay of speech through air and the reaction times of the interlocutors. If we assume the one way propagation time between two people A and B to be $t$ and the reaction time is $r$, the delay until a response from B to an utterance by A is returned to A, is

$$T_f = 2t + r$$

(i.e. the propagation delay of A's utterance plus B's reaction time plus the propagation delay of B's response back to A, see Figure 4a).

Similarly, for a telepresence conversation, the perceived response time is determined by the propagation time in air from the speaker's mouth to the microphone of the telepresence system, the combined processing and (electronic or optic) propagation delay of the speech signal through the telepresence system and the communication link, the propagation time in air from the telepresence system's speaker to the listener's ear, the reaction time of the responding participant and then the same propagation times back to the originator of the verbal interaction, as illustrated in Figure 4b. This gives us a complete delay of

$$T_t = d_1 + d_2 + p_1 + p_2 + d_3 + d_4 + r.$$

Since the microphone and speaker may be at different distances from the user, $d_1$ is not necessarily equal to $d_2$ in Figure 4b, and for the same reason $d_3$ is not necessarily equal to $d_4$. Furthermore, $d_1$ is not necessarily the same as $d_3$ and $d_2$ may be different from $d_4$. However, since we are only interested in aggregate delays we may assume without loss of generality that $d_1 = d_2 = d_3 = d_4 = d$. Similarly, we can assume $p_1 = p_2 = p$ although not strictly necessary for asymmetric communication links. This gives us a total delay for the telepresence situation of

$$T_t = 4d + 2p + r.$$

The difference in delay, $\Delta$, between the face-to-face situation and the telepresence situation is hence

$$\Delta = T_t - T_f = 4d + 2p + r - 2t - r = 4d + 2p - 2t.$$

In order for a telepresence system not to introduce any delay at all compared to the face-to-face situation (i.e. $\Delta \leq 0$), $p$ must be less than $t - 2d$, which means that the microphones and speakers of the telepresence system must be placed in a way that the propagation time of speech through air is smaller than in the face to face situation by a value bigger or equal to the combined processing and electric/optic transmission time of the telepresence system. To explore this implication in practice, we assume a 2 m face to face interlocutor distance, which gives t to be about 6 milliseconds, and the microphone-mouth and speaker-ear distances to be 0.1 m, which gives d to be about 0.3 ms and hence $p \leq 6 - 0.6 = 5.4$. Thus, we can tolerate a joint one-way processing and transmission delay of around 5 ms, without introducing any total delay, or equivalently about 10 ms round-trip delay. (This is the reason for the 10 ms delay value being used in section 5.1.5.)

Such a strict delay limit on a telepresence system might seem overly restrictive, since a small additional delay can be expected to go unnoticed. To explore how large additional delays can be tolerated, we first note that experimental studies on the effect on transmission delay in distributed conversations have usually focused on establishing when a delay becomes large enough to be annoying or detrimental to task performance, rather than to identify perceptual limits on detection of delays. Many such tests have been performed (e.g. [22]) and although inconclusive they suggest that delays of a few hundred milliseconds can be acceptable without significantly affecting user satisfaction or task performance. Moreover, conversation analysis in interactional linguistics suggests that the threshold for detecting between-speaker silences in conversation lies close to 200 ms [23]. Appreciating that it is exclusively in turn-taking that the effects of additional delay imposed by a telepresence system is detectable, we furthermore note that conversation analysis has shown that turn-taking gaps in conversations tend to vary from zero to about one second with a median around 200 ms. Overlaps between speakers are relatively common, i.e. double-talk where one person starts talking before the current speaker has finished. Overlaps also vary between 0 and one second, so the turn-taking interval can be seen to be between -1 second and 1 second relative to the end of an utterance, with a Gaussian distribution slightly skewed to the positive end, with approximately 60% of turn-taking gaps being positive [24].

Taken together, this data suggests that $\Delta \leq 200$ ms can be hypothesized to be good enough to pass unnoticeably and hence a target value for the maximum round-trip delay of a telepresence system.

### 5.4.2. Distance-spanning limits on telepresence

Since part of the delay imposed by a telepresence system is due to the finite speed of light, the delay requirement puts a hard upper limit on the geographical distance a telepresence system passing the TTT can span. If we assume $\Delta = 200$ ms, and the speed of light in a fiber cable to be 200,000 km/s, the theoretical maximum distance becomes 40,000 km, which incidentally is approximately the earth's circumference, allowing for earthbound telepresence installations to pass the TTT irrespective of geographic separation between the sites. If, on the other hand, we insist on $\Delta = 0$, the theoretical maximum distance a telepresence system passing the TTT can span is around 2000 km.

## 5.5. Other quality requirements

In addition to the requirements on audio and video quality and the requirements related to the effects of communication delay and jitter, there are also requirements on the design of the combined physical and virtual space created by the telepresence system. These requirements can be collectively described as the quality of immersion.

There are also many non-functional requirements related to the capacity or performance of the technological components of the system, such as the availability of sufficient network bandwidth.

### 5.5.1. Quality of Immersion

The quality of immersion must be high enough for remote users to be perceived as physically present, which means that the video-mediated or synthetic parts of the visual field of all users must blend perfectly with the physical parts. One can also regard the aural impressions as part of the immersive experience, so that the users are in fact immersed in a shared soundscape as well as in shared visual environment. The technology used by the system should be transparent, or at least sufficiently undetectable not to reveal that it is in fact mediating the communication session. Although technology for immersive virtual environments has improved considerably since the concept of telepresence was invented, the goal of achieving completely transparent technology immersing users in a shared space is still a very challenging topic not fully realized by state of the art systems.

## 6. CONCLUSION

I would now like to turn back to the question posed in the introduction, "Can technology eliminate the effect of distance in interpersonal communication?" and try to give an answer based on the reformulated question of whether a telepresence system can pass the TTT or not.

It is unquestionably the case that high-end telepresence systems available today can give its users a considerable feeling of presence. It is to my mind equally obvious that no system available today will fool anybody into believing that a remote user is physically present at the same location. Despite the tremendous advances in display, camera, multimedia processing and communication technologies as well as progress in algorithmics, software development and other relevant disciplines, state of the art telepresence systems will stand little or no chance of passing the TTT. As the analysis of the quality requirements of telepresence of this paper shows, many technological challenges still need to be overcome, particularly concerning multiview stereoscopic visualization, acoustic echo cancellation, low overall processing and communication latency as well as practical issues with immersion and transparency of technology.

The next question, then, is whether telepresence technology will ever be powerful enough to pass the TTT, and if so when. I strongly believe that in the near future, telepresence technology will evolve into a state where systems passing the TTT can be realized. This will require continued research and technology development within the fields of signal processing, immersion technologies, display and video capture technology, computer science and behavioural science.
To quote Turing, we can only see a short distance ahead, but we can see plenty there that needs to be done.

## AUTHOR

Mathias Johanson was born 1971 in Gothenburg, Sweden. He received his MSc degree in Computer Science in 1994 and his PhD degree in 2003 from Chalmers University of Technology. He is co-founder and R&D Manager of Alkit Communications AB. 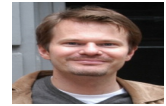